\documentclass[conference]{IEEEtran}
\IEEEoverridecommandlockouts

\ifCLASSINFOpdf

\else

\fi
\usepackage{float}
\usepackage{caption}
\usepackage{subcaption}

\usepackage{amsmath}
\usepackage{comment}

\usepackage{graphicx}
\usepackage{multirow}
\usepackage{svg}
\usepackage{caption}

\begin{document}
\title{Development of Hardware-in-Loop Framework for Satellite Communication Self-Healing Networks\\
 \thanks{This research work was supported by Lockheed Martin Space.}
} 

 \author{
    \IEEEauthorblockN{Sambrama\IEEEauthorrefmark{1}, Venkata Srirama Rohit Kantheti\IEEEauthorrefmark{1}, Liang C Chu\IEEEauthorrefmark{2}, Erik Blasch\IEEEauthorrefmark{3}, and 
    Shih-Chun Lin\IEEEauthorrefmark{1}}
    
    \IEEEauthorblockA{\IEEEauthorrefmark{1}North Carolina State University, Raleigh, NC, USA,
     \{sambram, vkanthe, slin23\}@ncsu.edu}
     
    \IEEEauthorblockA{\IEEEauthorrefmark{2}Lockheed Martin Space, Sunnyvale, CA, USA, 
    liang.c.chu@lmco.com}
    
    \IEEEauthorblockA{\IEEEauthorrefmark{3}Air Force Research Lab, Rome, NY, USA,
    erik.blasch.1@us.af.mil}
    
}

\maketitle

\begin{abstract}
The use of Low Earth Orbit (LEO) satellites in the next generation (Next-G) communication systems has been gaining traction over the last few years due to their potential for providing global connectivity with low latency. Since they are the closest to the earth they come with their own set of disadvantages including high vulnerability to jamming and interference. To address these issues, this paper introduces a resilient, self-healing network designed to optimize signal quality under dynamic interference and adversarial conditions. The network leverages inter-satellite communication and an intelligent algorithm selection process, incorporating combining techniques like distributed-Maximal Ratio Combining (d-MRC), distributed-Linear Minimum Mean Squared Error Estimation (d-LMMSE), and Selection Combining (SC). These algorithms are selected to improve performance by adapting to changing network conditions. To evaluate the effectiveness of the proposed solution, we develop a software-defined radio (SDR)-based hardware testbed and perform detailed performance evaluations. Additionally, we present results from field tests conducted on the AERPAW testbed, which validate the proposed combining solutions in real-world scenarios. The results show that our approach makes LEO satellite networks more reliable and better able to handle interference, making them suitable for critical communications.
\end{abstract}


\IEEEpeerreviewmaketitle

\section{Introduction}

The advancements in communication standards and system architectures have directly contributed to the rapid proliferation of satellites, majorly in the area of LEO satellite constellations. Satellite communications come in with the broad range of use cases, including, but not limited to media broadcasting and environmental monitoring. The growing interest in this area is largely influenced by the need for providing high quality internet to the remote areas, the ease of integration of various wireless technologies into the Next-G communication systems, and the rising demand for connected devices for applications such as smart cities and industrial automation \cite{satcom}. Recent years have seen the growth of using LEO satellites in a larger constellation network. 

LEO mega-constellations are a promising solution for providing global connectivity and bridging the digital divide. These distributed satellite swarms offer low-latency, overlapping coverage without geographical limitations, making them ideal for broadband connectivity. By working together with ground stations and autonomous systems, they can adjust their positions to maintain reliable and flexible service, supporting applications like IoT, M2M communication, and smart transportation in future networks \cite{leo5g}, \cite{6g}, \cite{sajid}. However, one major challenge is the strong line-of-sight (LOS) channel between ground terminals and satellite nodes, which limits channel diversity and hinders the implementation of SIMO and MIMO communications. To address this, distributed satellite swarms can introduce the necessary diversity \cite{infocom}. Additionally, such diversity enhances jamming resilience, as LEO satellites are vulnerable to attacks due to their fixed orbits and cyclic visibility. Anti-jamming techniques like spread spectrum methods are vital for ensuring secure satellite communication, and by leveraging diversity in swarm configurations, the impact of jammers can be minimized, improving the overall reliability of the network \cite{milcom}.

Self-healing is the ability of the system to detect and recover from abnormalities with minimal or no human intervention \cite{sh1}. 
They play an important role in complex environments like the ones involving satellites, where having manual intervention is not feasible. Self-healing networks help network function smoothly by handling issues including interference and node failures. We intend to build a self-healing network that intelligently switches between various combining techniques depending on the situation. This ensures that the system is reliable and has high performance even in adverse conditions. 

To ensure reliable communication between satellites in a LEO constellation, Inter-Satellite Links (ISLs) are typically used \cite{dynamicpoweralloc}. These links not only enhance anti-jamming performance but also improve the overall robustness of the satellite networks \cite{isl1}. By establishing direct communication channels, ISLs allow satellites to exchange data, collaborate with shared resources, and maintain reliable connections even in challenging environments, thus building an autonomous system with minimal human intervention. Therefore, we develop a system with device-to-device communication functionalities to mimic the characteristics of ISLs.

The proposed hardware end-to-end system is implemented using the Universal Software Radio Peripheral (USRP) B210's along with Raspberry Pi's (RPi) to enable self-healing on bitstream data. We also implement self-healing on images showing the capabilities of the algorithms for larger dataset. The d-MRC and the d-LMMSE algorithms used in self-healing along with the signal processing techniques used to process and extract the data have already been introduced in our previous work \cite{milcom}. The wireless transmission of the extracted data among the RPi's happens seamlessly through an internal ad-hoc network mimicking the ISL. Real world capabilities of the combining algorithms are validated using the data captured on the AERPAW testbed \cite{aerpaw}.

The remainder of the paper is divided as follows. Section II discusses the proposed system architecture and the indoor hardware testbed. Section III introduces the algorithms and the state machine used for our self-healing network. Section IV discusses about inter-device links. In section V, we present the results and we finally conclude the paper in section VI.

\section{System Architecture}
This section presents the proposed architecture along with the in-lab hardware testbed used to validate our algorithms.

\subsection{Proposed Architectural Model}
\begin{figure}
\centering
    \includegraphics[width=1\columnwidth]
    {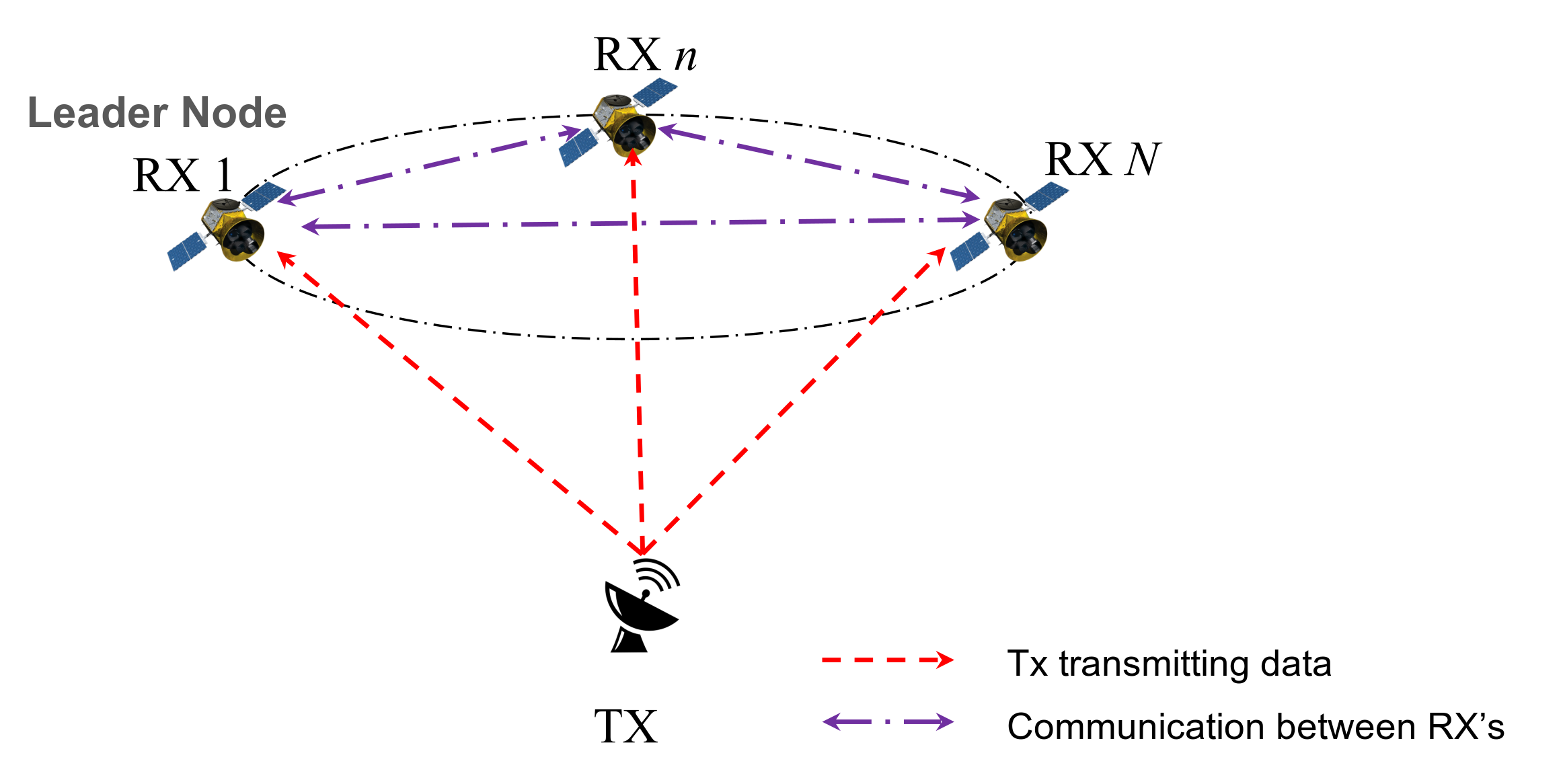}
    \captionsetup{justification=centering}
    \caption{Proposed architecture.}
    \label{fig:proposed_arch}
\end{figure}

Fig. \ref{fig:proposed_arch} illustrates the proposed system architecture, which consists of a Single-Input Multi-Output (SIMO) setup with one transmitter (TX) and $N$ receivers (RX). The transmitter represents a ground terminal that continuously transmits signals to the swarm of receivers positioned in Low Earth Orbit (LEO). These receivers, which are part of a satellite swarm, are connected through an internal ad-hoc network that enables device-to-device communication and coordination. This network allows each satellite to share locally processed signal data with the rest of the swarm. A designated leader node within the swarm is responsible for collecting the processed data from all RX units and applying the appropriate combining algorithm to enhance signal quality. To ensure robustness and avoid single points of failure, the leader node role is dynamically assigned to different RXs in a sequential manner. In the example shown in the figure, RX$1$ is acting as the leader node at that particular instance, demonstrating the flexibility and distributed control of the system.

\subsection{Experimental Testbed Setup}
\begin{figure}
\centering
    \includegraphics[width=1\columnwidth]{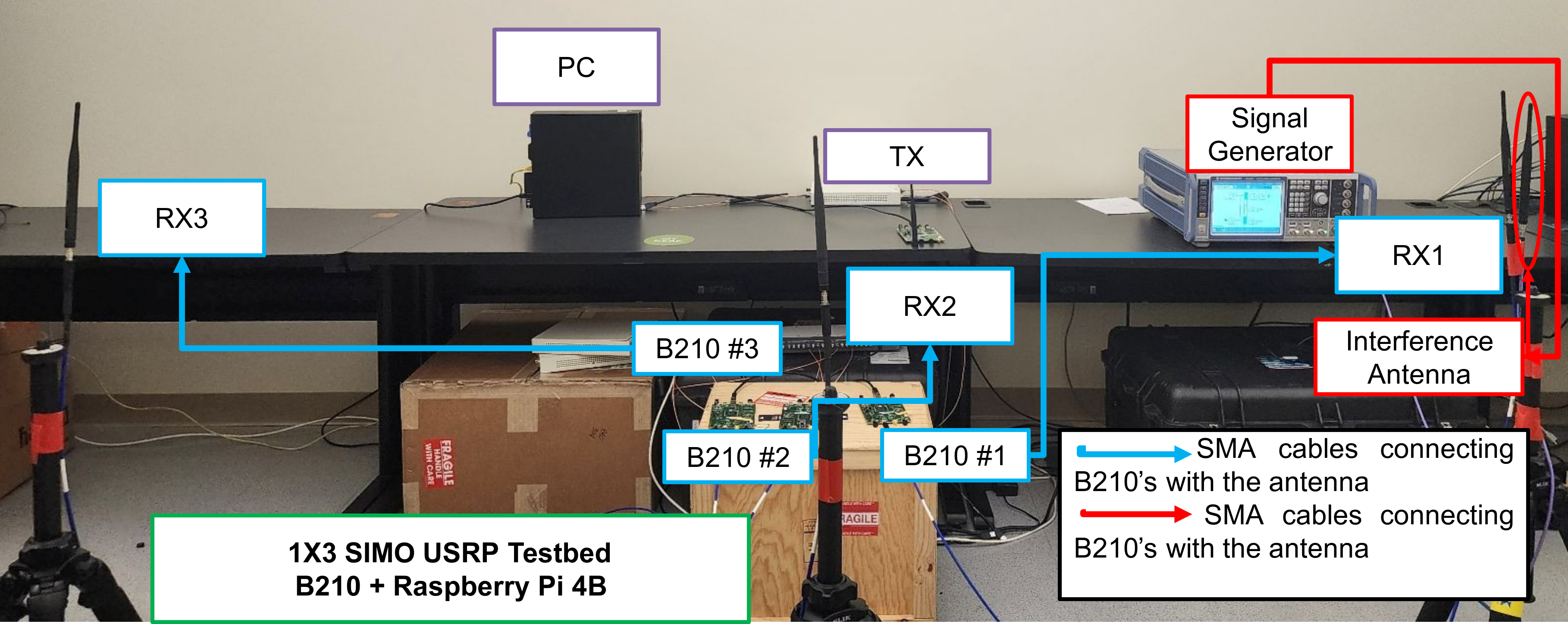}
    \captionsetup{justification=centering}
    \caption{Indoor testbed for $1\times3$ SIMO testbed.}
    \label{fig:b210_testbed}
\end{figure}

Fig. \ref{fig:b210_testbed} shows the testbed setup used to implement the system described above. The setup includes three USRP B210 software-defined radios, each connected to a RPi, functioning as the receivers. An additional USRP B210 is connected to a PC and serves as the transmitter (TX). The RPis control the RX units, and their compact size and low power requirements make them highly portable, enabling a flexible and mobile testing environment. An antenna connected to a signal generator is placed close to RX$1$ to introduce jamming into the system. The USRP B210s support two wireless channels, allowing them to function as either single-antenna or multi-antenna devices depending on the use case. For this SIMO-based implementation, each USRP is configured to operate as a single-antenna radio.

\section{Self-healing Proposal}
This section presents the algorithms included in our setup, their results for standalone implementation and the proposed self-healing network.

\subsection{Algorithms Implemented in Self-Healing}
We transition among three algorithms to obtain superior quality signal. The algorithms have already been introduced in \cite{milcom} and are explained in brief here.

\subsubsection{Distributed Maximal Ratio Combining}
The d-MRC algorithm utilizes a weighted combination of data sequence of different receivers, contingent on individual channel quality.

The signal data at each of the RX can be written as,
\begin{align}
\mathbf{r}_i[k]= \sqrt{p_g}h_i\mathbf{x}[k] + \sqrt{p_j}h_{j,i}\mathbf{x_j}[k] + \mathbf{n}_i, 
\label{dt_rcvd_data_siso}
\end{align}
where, $p_g$ and $p_j$ are the transmitter and jammer transmit power respectively, $h_i$ and $h_j,i$ are channel information for the $i$th channel, $\mathbf{x}$ and $\mathbf{x_j}$ are the input sequences for the transmitter and the jammer respectively and $\mathbf{n}_i$ is the iid noise. 

In accordance with Eq. \eqref{dt_rcvd_data_siso}, the data from all the receivers are collected into a single data structure for further processing. If there are a total of $N$ receivers, then $\mathbf{R} = [r_1,r_2,...,r_N]^T$, represents this data structure. d-MRC exploits uncorrelated channels in wireless communication to combine signals that carry the same information. It multiplies the data at all receivers with a weight vector $\mathbf{u} =[u_1,u_2,...,u_N]^T$, such that the effective SNR is maximized. The interference is assumed to be negligible and thus $\sqrt{p_j}h_{j,i}\mathbf{x_j}[k]$ can be ignored. The SNR for the combined result can thus be written as,
\begin{align}
\mathrm{SNR} = \frac{{p_g}{\sum^{N}_{(i=1)}{||u_i^Hh_i||^2}}}
{\sigma^2{\sum^{N}_{(i=1)}||u_i||^2}},
\label{mrc}
\end{align}
where $\sigma^2$ is the AWGN noise power level, $\mathbf{h} = [h_1,h_2,...,h_N]^T$ represents a vector of channel state information from $N$ receivers and $(.)^H$ is the Hermitian operator.

On optimizing as per \cite{heath2017introduction} and after applying the Cauchy-Schwarz Inequality, we obtain the d-MRC weight $\mathbf{u}_{d-MRC}$ as,
\begin{align}
\mathbf{u}_{\mathrm{d-MRC}} = \mathbf{h}^H.
\end{align}

\subsubsection{Distributed Linear Minimum Mean Square Error Estimation}
The d-LMMSE algorithm also utilizes a weighted combination of data sequence but includes the interference data and thus, is useful for signals with high levels of interference such as jamming. The effective SINR can be written as,
\begin{align}
\mathrm{SINR} = \frac{{p_g}{\sum^{N}_{(i=1)}{||u_i^Hh_i||^2}}}
{{\sigma^2\sum^{N}_{(i=1)}||u_i^H ||^2} +{{p_j}\sum^{N}_{(i=1)}||u_i^Hh_{j,i} ||^2}}.
\end{align}
Optimal value of $\mathbf{u}$ is obtained by maximizing SINR by solving the above equation as a minimum mean squared error optimization problem. By assuming that the transmitted signals are legitimate and thus, external interference signals and additive noise are mutually independent, we obtain, 
\begin{align}
\mathbf{u}_{d-LMMSE} = {p_g}[{p_g}\mathbf{h}\mathbf{h}^H + {p_j}\mathbf{h}_j\mathbf{h}_j^H + \sigma^2\mathbf{I}]^{-1}\mathbf{h},
\end{align}
where, $\mathbf{I}$ is the identity matrix of size $N \times N$.

\subsubsection{Selection Combining}
Selection Combining selects the best among the $N$ receivers at any given point of time \cite{seldiv}. 
Mathematically, this will be similar to Eq. \eqref{mrc}, where the RX's that are highly impacted with interference will have a weight vector of $0$. This leaves the combined signal to be equal to the signal with legitimate data, thus making the weight vector for that particular RX to be $1$.

\subsection{Results for Standalone Algorithm Implementation}
The algorithms discussed in the previous section were tested individually to assess how well they would work within the self-healing framework. These experiments were conducted in an indoor laboratory setting using a $1\times2$ SIMO testbed using USRP X310's for various transmit powers with a constant interference of $-115$ dBm placed close to RX$1$. The results are presented in Table \ref{tab:table1}. We observe that d-MRC performs well for higher transmit power, when both the RXs have higher datarates. d-LMMSE is the best performing algorithm when one of the RXs has been affected, while SC performs the best for low power levels with both the RXs getting affected. This proves the need for using self-healing to transition between the three algorithms in order to have superior quality signals in diverse environments.



\begin{table}[b]
\caption{Standalone Implementation Results}
\centering
\begin{tabular}{|c|ccccl|}
\hline
\multirow{2}{*}{\textbf{\begin{tabular}[c]{@{}c@{}}Transmit Power \\ (dBm)\end{tabular}}} & \multicolumn{5}{c|}{\textbf{Datarates (Kbps)}} \\ \cline{2-6} 
 & \multicolumn{1}{c|}{\textbf{RX1}} & \multicolumn{1}{c|}{\textbf{RX2}} & \multicolumn{1}{c|}{\textbf{d-MRC}} &
 \multicolumn{1}{c|}{\textbf{d-LMMSE}} &
 \textbf{SC} \\ \hline
-23.5 & \multicolumn{1}{c|}{410.01} & \multicolumn{1}{c|}{429.54} & \multicolumn{1}{c|}{465.60} & \multicolumn{1}{c|}{456.87} & \multicolumn{1}{c|}{429.54}\\ \hline
-35.5 & \multicolumn{1}{c|}{337.03} & \multicolumn{1}{c|}{427.52} & \multicolumn{1}{c|}{395.91} & \multicolumn{1}{c|}{441.86} &\multicolumn{1}{c|}{427.52} \\ \hline
-47.5 & \multicolumn{1}{c|}{253.51} & \multicolumn{1}{c|}{327.66} & \multicolumn{1}{c|}{326.69} & \multicolumn{1}{c|}{327.60} & \multicolumn{1}{c|}{327.66}\\ \hline
\end{tabular}%
\label{tab:table1}
\end{table}

\subsection{State Machine Model for Self-Healing}
\begin{figure}
\centering
    \includegraphics[width=0.8\columnwidth]{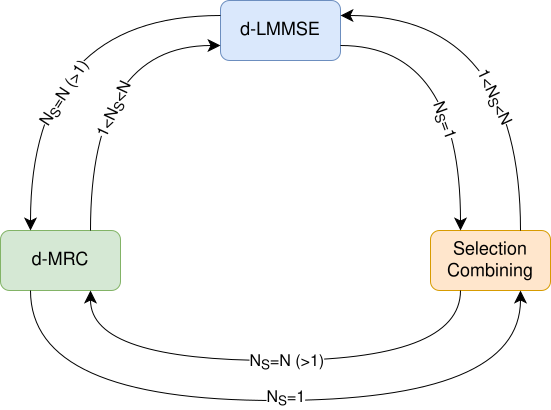}
    \captionsetup{justification=centering}
    \caption{State machine for self-healing.}
    \label{fig:state_machine_phy}
\end{figure}

Self-healing is implemented to make the system independent and have the best possible combining algorithm implemented at all times inorder to obtain superior quality signal. Fig. \ref{fig:state_machine_phy} represents the state machine used to implement the combining method.  The algorithm to be implemented is selected based on the number of receivers, $Ns$, having bit error rates (BER) greater than the threshold. One among d-MRC, d-LMMSE and SC gets chosen based on the BER's. $N$ represents the total number of RX's.

The figure clearly describes the implementation of the three combining algorithms. When all the RXs receive high quality signal, the BER for each of the RX units will be greater than the given threshold. In this case the number of receivers with BER greater than the threshold, $N_S$, will be equal to the total number of available receivers, i.e, $N_S=N$, prompting the system to implement d-MRC. In cases where the interference levels are extremely high and only one of the receivers receive good signal, i.e, $N_S=1$, leading the system to implement SC, where the best performing receiver is chosen. Another scenario is where the number of receivers receiving high quality signal is more than one but less than the total number of available receivers. Here, the BER is greater than the threshold for $1<N_S<N$ and thus d-LMMSE gets implemented.


\section{Inter-Device Links}

To develop a self-sufficient swarm system with ISL capabilities, we designed a network of RXs capable of direct device-to-device communication. Each RX shares data over a dedicated wireless channel, which is separate from the channel used for SIMO communication. This architecture allows the RXs to perform signal combining locally, removing the need for an external computer and making the swarm system autonomous.

\begin{figure}
\centering  
    \includegraphics[width=0.45\columnwidth]{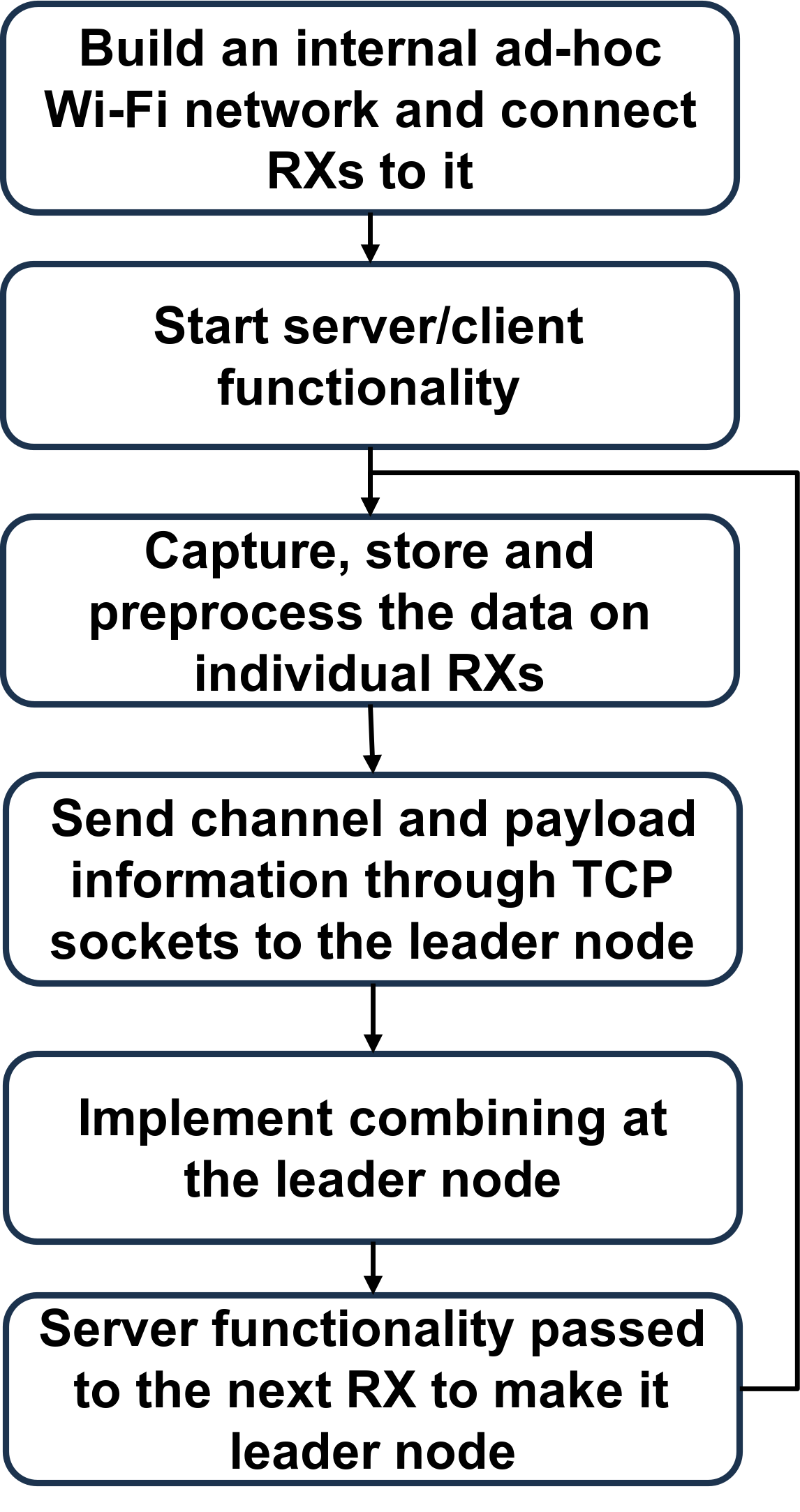}
    \captionsetup{justification=centering}
    \caption{Flow diagram for inter-device links.}
    \label{fig:line_diagram}
\end{figure}

Device-to-device communication is implemented using TCP sockets, where each RX can function both as a server and as a client. The flow diagram in Fig. \ref{fig:line_diagram} represents the flow of the architecture. Upon startup, all devices connect to a common Wi-Fi network using the onboard network interface cards on the RPis. Once connected, the RXs establish TCP server-client connections in a dynamic, rotating fashion \cite{dyspan}. Initially, RX$1$ assumes the role of the server, acting as the leader node, while the other RXs connect as clients. After a processing cycle, the server responsibility shifts to RX$2$, then to RX$3$, and so on, ensuring that each device takes turns coordinating the data exchange and processing.

During operation, each RX captures and processes its own signal data, and stores both channel and payload information locally. The client RXs then transmit their processed data to the current leader node via the established TCP sockets. At the leader node, the system computes the bit error rate (BER) for each receiver and compares these values to a predefined threshold. Based on this analysis, the leader node selects and implements the most suitable combining algorithm to optimize performance. Once the combining process is complete, leader node capabilities are passed to the next RX in the sequence, and the process repeats, enabling continuous and distributed operation without external oversight.

To enhance robustness, the RXs are programmed to automatically detect and recover from communication failures. If the connection is lost, the devices attempt to reconnect without requiring the system to restart. Should a particular RX become unavailable, the system simply excludes it from the combining process and continues operating with the remaining devices. This fault-tolerant design ensures the swarm remains resilient and independent of any single node.

\section{Over the air Implementation}
This section discusses the results obtained for self-healing network in the in-lab testbed. We present the results for transmitting bitstream and image data. We also present data obtained from real world testing using the AERPAW platform.

\subsection{Self-Healing on Bitstream Data}
The system is first tested for bitstream data, where a random set of data is generated using MATLAB. This data is modulated using QPSK modulation and then, appended with Zadoff-Chu preamble sequence. We use a centre frequency of $2.55$ GHz and a bandwidth of $1$ MHz. The data is continuously transmitted from the transmitter. The entire reception is automated to work with self-healing. The individual B210s capture this data, save it on the RPi which also performs signal processing including timing, frequency and phase compensation, determines the channel estimates and the payload information. The channel estimates and payload information are shared with the leader node through the inter-device links to implement the combining algorithm and obtain the datarates.

\begin{figure}
\centering
    \includegraphics[width=0.95\columnwidth]{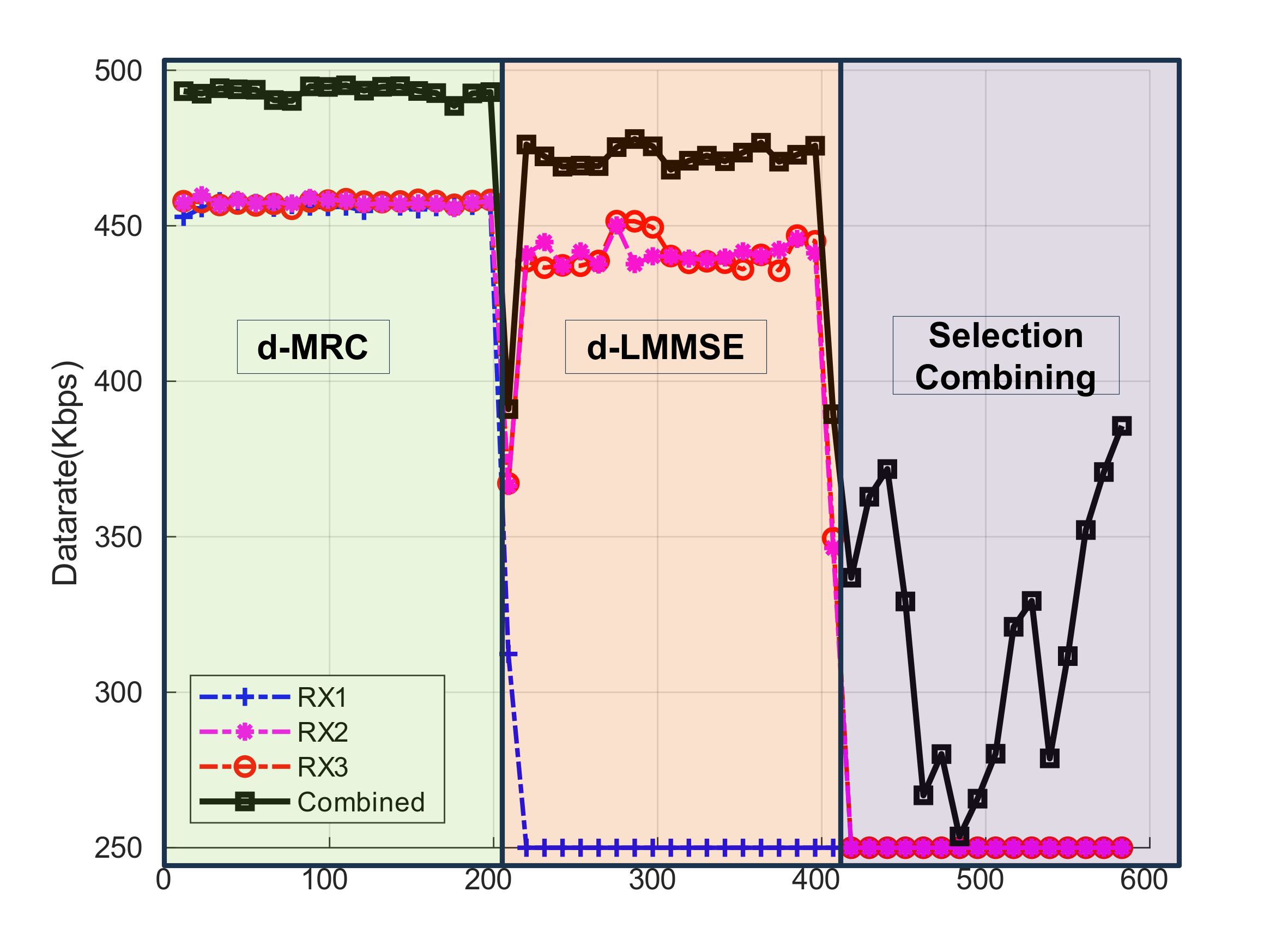}
    \captionsetup{justification=centering}
    \caption{Datarates for bitstream data.}
    \label{fig:bitstream_graph}
\end{figure}

Fig. \ref{fig:bitstream_graph} represents the continuous plot for bitstream data through self-healing implemented using the automated B210 setup. It shows that the combined output is always better than the individual datarate values validating our claim. At the points where selection combining is implemented, we observe that the combined output is equal to the datarate of the RX with good reception. We have induced a constant external interference with power $-20$ dBm using a signal generator and placed the antenna close to RX$1$. We start the experiment with high transmit power of $10.5$ dBm and observe that all three receivers have similar performance and have high datarates, which suggests that the d-MRC algorithm was implemented since the BER for all receivers exceeded the threshold, as specified by the state machine. We later decrease the transmit power to $-9.5$ dBm and notice that the receiver closest to the interference antenna, i.e, RX$1$ is impacted and doesn't receive good data, thus d-LMMSE is implemented. For the last part of this experiment, we further decrease the transmit power to $-29.5$ dBm and see that both RX$1$ and RX$2$ are highly impacted, RX$3$ is not impacted as much as the other two and thus selection combining is implemented and the combined datarate is equal to the datarate of the best performing receiver, which in this case is RX$3$.

\subsection{Self-Healing on Image Data}
With the self-healing working well for bitstream data, we look further to implement it on larger datasets and thus implement self-healing on image data. We make use of a small airplane figure as the signal to be transmitted. The image is converted to grayscale to reduce the overall size of the signal and is then vectorized and converted into bitstream before transmitting it, similar to the bitstream data transmission. The data received is processed similar to the bitstream data and the final images are reconstructed and displayed along with the datarates. 

\begin{figure}
\centering
    \includegraphics[width=0.95\columnwidth]{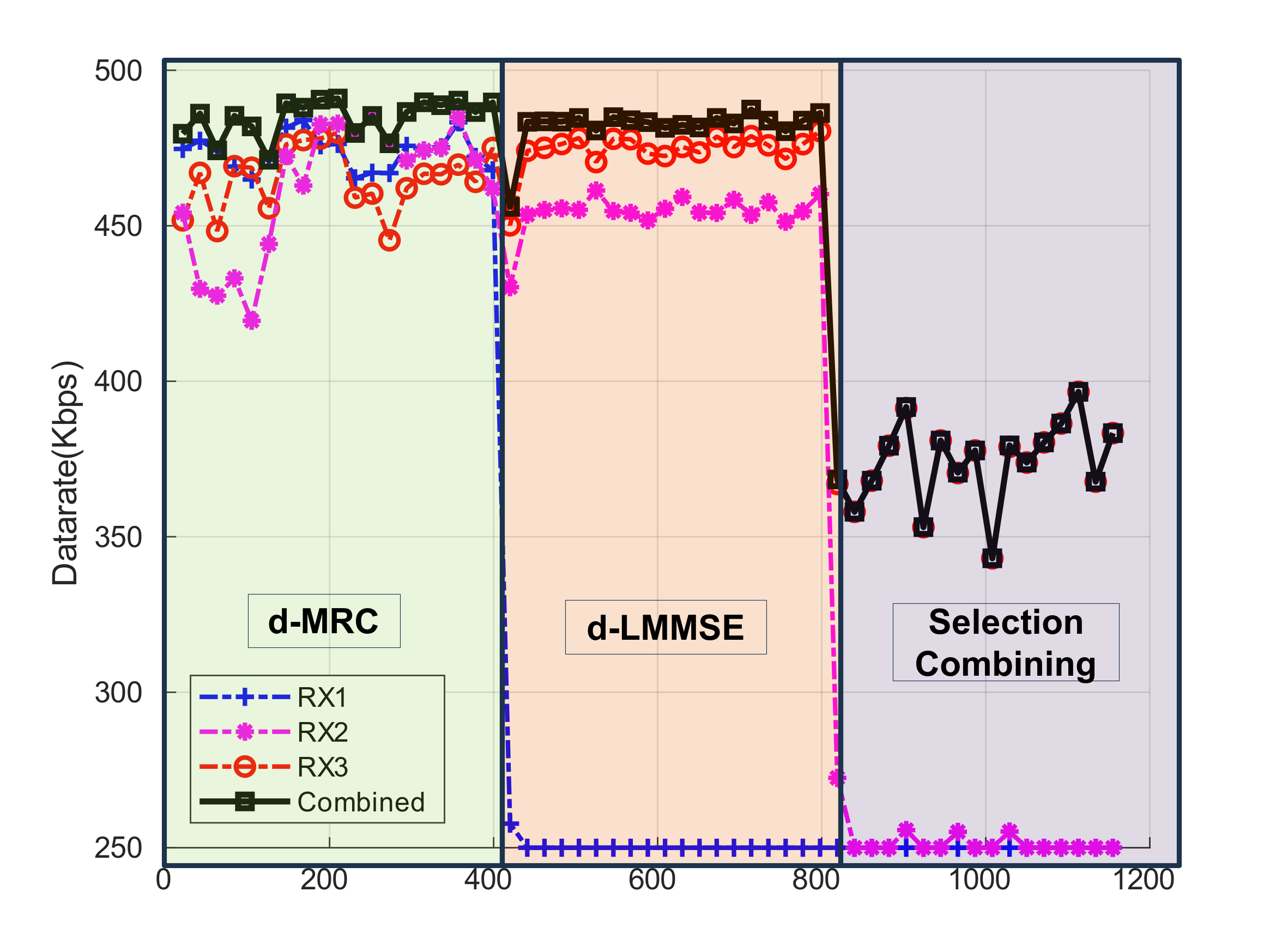}
    \captionsetup{justification=centering}
    \caption{Datarates for image data.}
    \label{fig:image_graph}
\end{figure}

\begin{figure*}
    \centering
    \begin{subfigure}{0.32\textwidth}
    \centering
    \includegraphics[width=\textwidth]{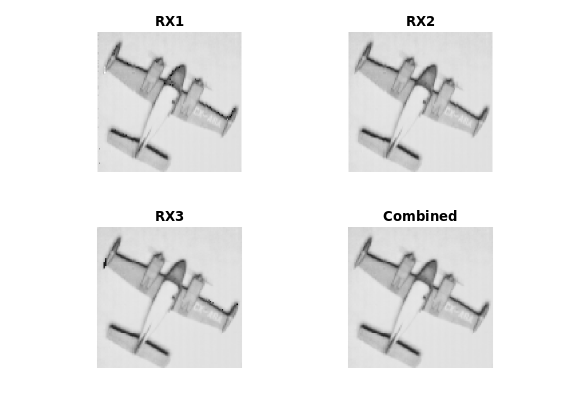}
    \captionsetup{justification=centering}
    \caption{Image for d-MRC Combining}
    \label{fig:mrc_image}
\end{subfigure}
     \centering
      \begin{subfigure}{0.32\textwidth}
\centering
\includegraphics[width=\textwidth]{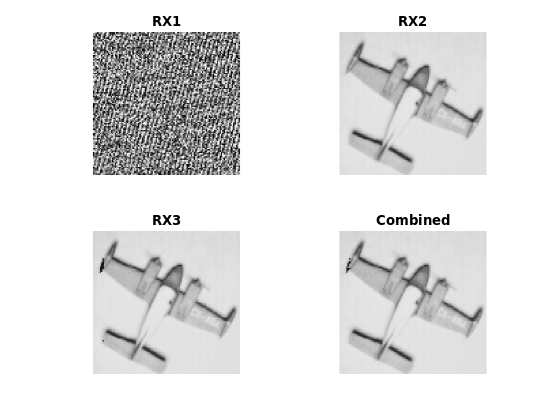}
    \captionsetup{justification=centering}
    \caption{Image for d-LMMSE Combining}
    \label{fig:lmmse_image}
 \end{subfigure}
 \begin{subfigure}{0.32\textwidth}
    \centering
    \includegraphics[width=\textwidth]{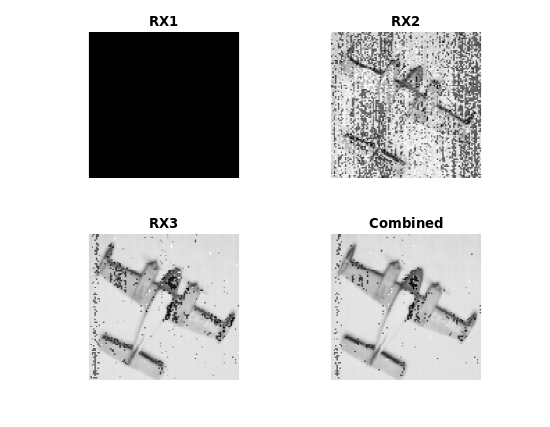}
    \captionsetup{justification=centering}
    \caption{Image for Selection Combining}
    \label{fig:siso_image}
    \end{subfigure}
     \captionsetup{justification=centering}
    \caption{Received Images}
\end{figure*}

Fig. \ref{fig:image_graph} represents the output for image transmission through B210's. Similar to the bitstream case, we see that the combined datarate is always better than the individual receivers, proving our algorithms work well even for larger datasets like images. The displayed images also demonstrate that not only is the datarate improved, but the overall quality of the received image is enhanced as well. We have induced a constant external interference with power $-30$ dBm using a signal generator. The antenna is placed close to the RX1 antenna. We start with the transmit power of $10.5$ dBm and see that all three receivers have similar performance and thus the d-MRC algorithm gets implemented. We later reduce the transmit power to $-7.5$ dBm and observe that the RX$1$ which is closest to the external interference antenna is heavily impacted and has no useful data received, thus d-LMMSE algorithm is implemented, with only RX$2$ and RX$3$ taken into consideration. We further reduce the transmit power to $-24.5$ dBm and notice that the two receivers closer to the external interference antenna are heavily impacted, leaving only RX$3$ to receive valid data and have higher BER than the threshold, thus SC gets implemented with datarate matching the datarate of RX$3$. 

Fig. \ref{fig:mrc_image} - Fig. \ref{fig:siso_image} are the sample reconstructed images at the receivers, displayed for each of the three cases. The figures clearly illustrate the working of our algorithms. Fig. \ref{fig:mrc_image} is the output for d-MRC implementation with all the RX's receiving high quality data. Fig. \ref{fig:lmmse_image} is the output for d-LMMSE implementation and it is clearly visible that RX$1$ is highly impacted and has not received any useful information and the combined image is the improvised version of RX$2$ and RX$3$.  Fig. \ref{fig:siso_image} illustrates the output for SC with RX$1$ and RX$2$ being heavily impacted and RX$3$ is also slightly impacted by the external interference. The combined image improvises on the best RX, which here is RX$3$.

\subsection{Field Test using AERPAW}

\begin{figure*}
\centering
     \begin{subfigure}{0.32\textwidth}
    \centering
      \includegraphics[width=\textwidth]{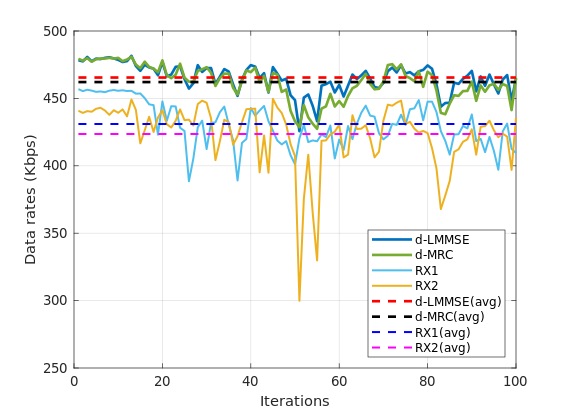}
    \captionsetup{justification=centering}
    \caption{UAVs taking different paths}
    \label{fig:Datarate - UAVs taking different paths}
    \end{subfigure}
    \centering
    \begin{subfigure}{0.32\textwidth}
    \centering
    \includegraphics[width=\textwidth]{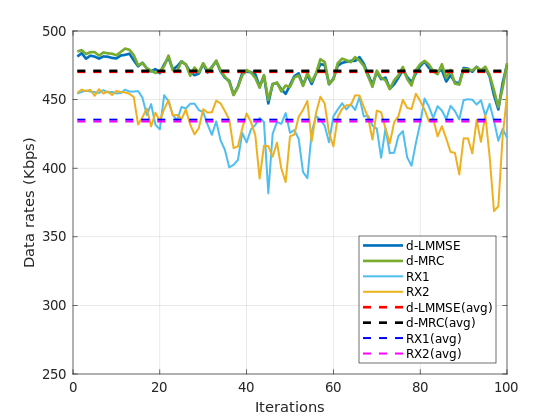}
    \captionsetup{justification=centering}
    \caption{Both UAVs taking path1}
    \label{fig:Datarate - Both UAVs taking path1}
    \end{subfigure}
      \begin{subfigure}{0.32\textwidth}
    \centering
    \includegraphics[width=\textwidth]{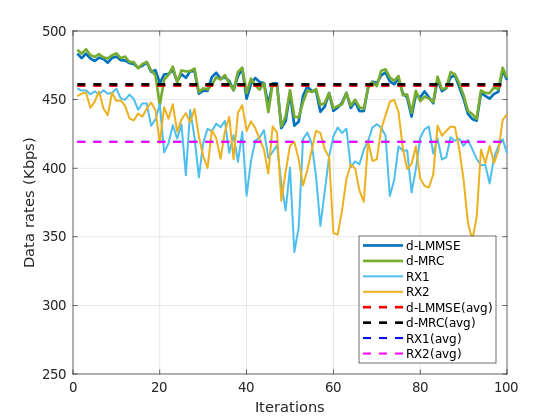}
    \captionsetup{justification=centering}
    \caption{Both UAVs taking path2}
    \label{fig:Datarate - Both UAVs taking path2}
    
    \end{subfigure}
     \captionsetup{justification=centering}
    \caption{Datarates for individual radios, d-MRC and d-LMMSE}
\end{figure*}

\begin{figure}
\centering
    \includegraphics[width=0.75\columnwidth]{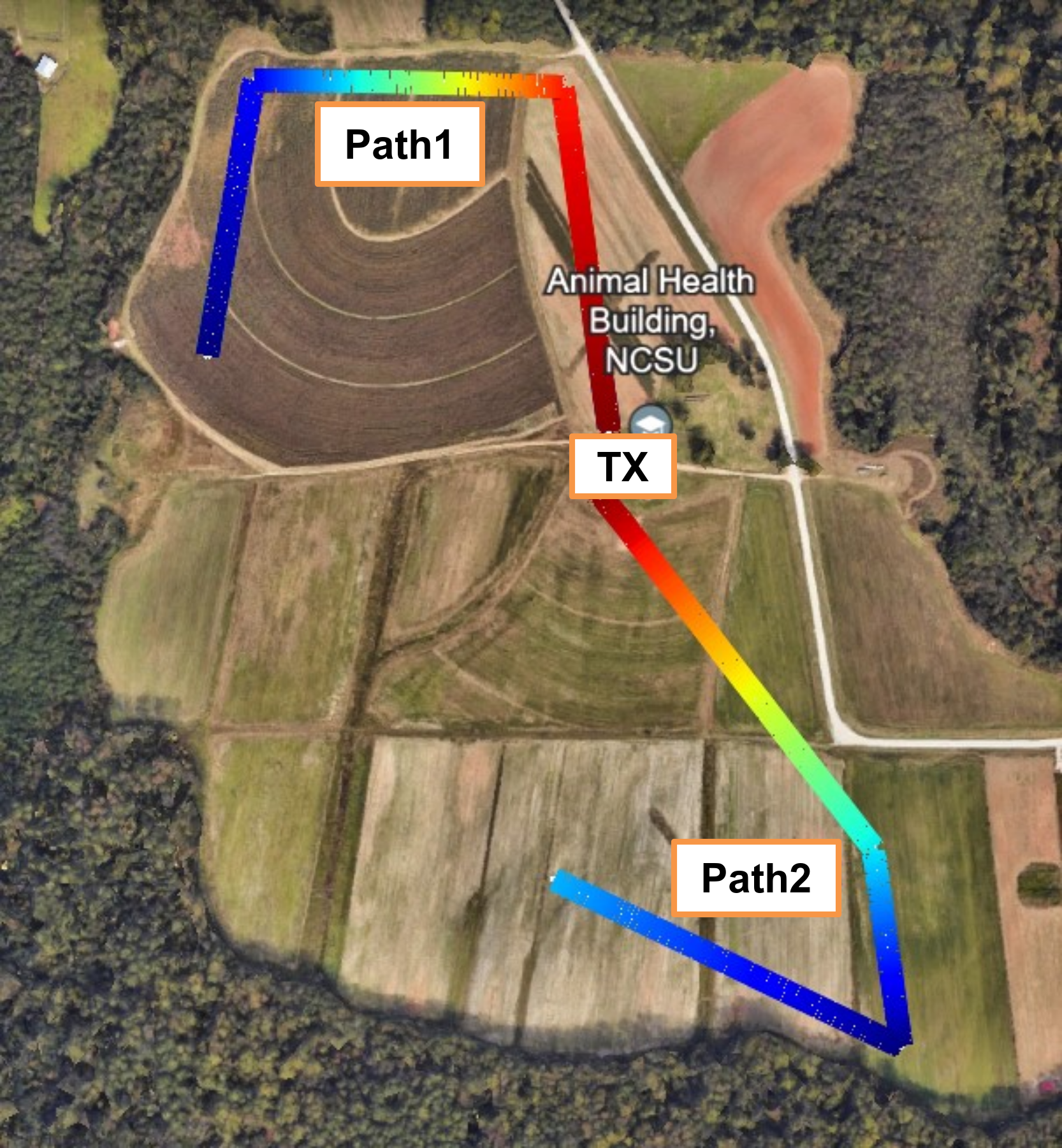}
    \captionsetup{justification=centering}
    \caption{Path for AERPAW experiments.}
    \label{fig:aerpaw_path}
\end{figure}

The AERPAW testbed at NC State enables experimental research in the real-world environment. We use the AERPAW testbed to transmit and receive data, which is then processed offline using our algorithms to ensure that our system is capable of providing good results when implemented in the real outdoor environment. The use of UAVs provides us with the environments close to that of LEO satellites. The experiment is performed on a $1\times2$ SIMO setup with two moving UAVs acting as receivers and a fixed transmitter. Fig. \ref{fig:aerpaw_path} shows the path taken by the UAVs.  We collect three sets of data with the first one being for UAV$1$ taking path$1$ and UAV$2$ taking path$2$. The second set of data is taken when both UAVs take path$1$ and the third set is for both the UAVs moving in path$2$. For all the $3$ cases, UAV$1$ is at a height of $50$ m and UAV$2$ is at $60$ m. Both the UAVs move at a speed of $5$ m/s and have a $25$ dB gain. The TX is fixed at a height of $9.144$ m and has a gain of $72.5$ dB. The experiments are conducted with a center frequency of $3.32$ GHz and a bandwidth of $1$ MHz.


Fig. \ref{fig:Datarate - UAVs taking different paths} - \ref{fig:Datarate - Both UAVs taking path2} shows the individual datarates, d-MRC and d-LMMSE results for all the three cases described above. It can be seen that the combined results are better than the individual datarates. We can also observe that if either of the receivers are extremely bad then d-LMMSE overperforms d-MRC. Each point in x-axis corresponds to the average datarate for $500$ messages and y-axis is the datarates in Kbps. The dotted lines are the average over the entire range.

\section{Conclusion}
This paper introduces a self-healing network architecture transitioning between the d-MRC, d-LMMSE and SC algorithms to obtain superior datarate values compared to the individual receivers are achieved across diverse environments in distributed SIMO systems. We evaluate the performance on bitstream and image data and show that our self-healing algorithm has superior performance in varied kind of environments. We effectively use device-to-device communication to share data among the devices and prove our ability in having ISL capabilities. We then validate the working of our algorithms in the real-world environment using the AERPAW testbed and show that the combined datarates are superior than the individual datarates. 

\ifCLASSOPTIONcaptionsoff
  \newpage
\fi

\bibliographystyle{IEEEtran}

\end{document}